# Over 30,000-fold field enhancement of terahertz nanoresonators enabled by rapid inverse design

*Hyoung-Taek Lee[1], Jeonghoon Kim[1], Joon Sue Lee[2], Mina Yoon[3], and Hyeong-Ryeol Park[1,*]*

[1]Department of Physics, Ulsan National Institute of Science and Technology (UNIST), Ulsan 44919, South Korea

[2]Department of Physics and Astronomy, University of Tennessee, Knoxville, Tennessee 37996, USA

[3]Materials Science and Technology Division, Oak Ridge National Laboratory, Oak Ridge, Tennessee 37831, USA

*nano@unist.ac.kr

ABSTRACT

The rapid development of 6G communications using terahertz (THz) electromagnetic waves has created a demand for highly sensitive THz nanoresonators capable of detecting these waves. Among the potential candidates, THz nanogap loop arrays show promising characteristics but require significant computational resources for accurate simulation. This requirement arises because their unit cells are 10 times smaller than millimeter wavelengths, with nanogap regions that are 1,000,000 times smaller. To address this challenge, we propose a rapid inverse design method using physics-informed machine learning, employing double deep Q-learning with an analytical model of the THz nanogap loop array. In about 39 hours on a middle-level personal computer, our approach identifies the optimal structure through 200,000 iterations, achieving experimental electric field enhancement of 32,000 at 0.2 THz, 300 % stronger than prior results. Our analytical model-based approach significantly reduces computational resources, offering a practical alternative to numerical simulation-based inverse design.

TEXT

Introduction

Following the development of terahertz (THz) electromagnetic wave generation and detection, numerous studies have been conducted in the field of materials characterizations as well as THz photonic applications over the past few decades[1-5]. The gigahertz to terahertz region is of great significance not only for the current 5G communication technologies, but also for the next generation of 6G and beyond[6]. In 6G technology, the frequency ranges between 75 and 300 GHz (0.075 to 0.3 THz) are being considered, posing challenges in developing high-performance detectors that operate at room temperature and cover large areas[7]. To address the need for significantly improved THz detector performance, the gap plasmon effect can be exploited to achieve strong electric field enhancement with a nanometer-scale metal-insulator-metal structure[3]. By leveraging this effect, the absorption cross section of materials can be improved through enhanced light-matter interactions, enabling a wide range of applications, such as high-performance detectors and ultra-low-density material sensing[8-12]. In line with these observations, many studies have explored terahertz nanoresonators, such as nano-split ring resonator (SRR)[13], slot antenna[14], flexible nanogap[15], and nanogap loop[16, 17]. Previous studies have demonstrated field enhancements of approximately 10,000 times at one of the 6G communication frequencies, 140 GHz, using 2 nm gap loop arrays, as well as 25,000 times at 75 GHz[16]. This remarkable field enhancement in combination with phase transition materials, such as $VO_2$ and superconductors[18-20], whose optical properties change with temperature, suggests the possibility of ultrasensitive detectors operating at 6G communication frequencies.

Designing practical photonic device by exhaustively exploring all parameter possibilities, fabricating, and evaluating each device is a time-consuming and costly process. Therefore, numerical simulations such as finite-difference time-domain (FDTD) and finite element method (FEM) have become essential tools for device design. Recently, there has been a significant increase in researchers adopting the inverse design approach, which utilizes artificial intelligence (AI) techniques to streamline the design process and achieve optimized designs[21-23]. This methodology has found applications in many fields, enabling the prediction of both intuitive and non-intuitive designs tailed to specific applications[24-38].

Inverse design typically involves tens of thousands of iterations. Previous research in this area has primarily focused on simulating structures that are 1 to 100 times smaller than the wavelength, as shown in Figure 1a. While the combination of numerical simulation and inverse design offers advantages for a variety of structures, it typically takes more than a month to achieve an optimal structural design[30]. For example, a three-dimensional numerical simulation for a specific state takes more than minutes, and the inverse design method may require 30,000 iterations for convergence. In particular, for THz nanogap loop arrays, as depicted in Figure 1b, the unit cells are ten times smaller than the wavelength, and nanogap regions are a million times smaller. In this case, one simulation with a specific THz nanogap loop array can take several tens of hours and require over 100 GB of RAM. Consequently, simulations for inverse design are considered impractical in this context. To address this challenge, we propose an inverse design method based on analytical solutions specifically tailored for terahertz nanogap loop arrays, aiming to achieve higher electric field enhancement than previously achieved[16]. We successfully obtained the optimal design within a reasonable time of about 39 hours on a typical middle-level PC (CPU: 3.40 GHz, 6 cores, 12 threads, RAM: 16 GB, GPU: NVIDIA GeForce GTX 1050), and further validated the results through simulations. In addition, large-area THz nanogap loop array structures were fabricated and experimentally analyzed using terahertz time-domain spectroscopy (THz-TDS).

## Results and discussion

**Physics-informed machine learning for inverse design of terahertz nanogap loop arrays.**

The term "physics informed machine learning" is commonly used when incorporating physical knowledge into machine learning algorithms to enhance their convergence[39, 40]. By incorporating theoretical models, such as the RLC model for the split-ring resonators[41], the modal expansion for slot antennas[42], or the Galerkin method for nanogap loop structures[43], into the inverse design process, computational resources and time required for design optimization can be significantly reduced. In the case of THz nanogap loop arrays, where metals are considered as perfect electric conductors (PEC) in the THz range due to their high optical conductivity, analytical solutions can be obtained for structures consisting of simple components, such as periodically arranged cylindrical or rectangular hole arrays[9, 17, 44]. The analytical solution for our nanogap loop array structure can be found in the Supporting information S1. Utilizing this method, the computation time for calculating the transmission spectrum is approximately 0.36 seconds per iteration on a typical middle-level PC (CPU: 3.40 GHz, 6 cores, 12 threads, RAM: 16 GB, GPU: NVIDIA GeForce GTX 1050), while the computation time for simulating with FEM requires approximately 20 hours to calculate the same spectrum. Therefore, by combining the analytical solution with the inverse design method, not only is the computational cost significantly reduced compared to simulation methods, but the design time is also reduced by approximately 480,000 times. The double deep Q-learning Environment and Agent for the inverse design of the THz nanogap loop arrays are configured as shown in Figure 2a. Here, to prevent the overestimation of Action values and mitigate interference with learning, the double deep Q-learning algorithm is used instead of the deep Q-learning in the reinforcement learning algorithm[22, 45]. The input state consists of six parameters: a width of $a_x$, a height of $a_y$, a horizontal period of $l_x$, a vertical period of $l_y$, a film thickness of h, and a gap width of w of the rectangular pattern, all of which influence the spectral features. There are a total of 13 Actions, with 12 Actions corresponding to the increase or decrease of six parameters, and one Action representing no change. The process of obtaining the Reward in our inverse design method in detail is as follows:

When an x-polarized light is incident on the nanogap loop structure, the analytical calculation of the transmitted electric field amplitude, as shown in the Supporting information S1, yields the electric field enhancement. The electric field is funneled into the insulating gap material, resulting in the huge enhancement at the gap[3]. Experimentally, the time transient obtained through THz-TDS can be Fourier transformed into a frequency domain spectrum. By conducting experiments on the sample and reference, we can obtain the far-field transmitted electric field amplitude, called the transmitted amplitude, as follows:

$$t = E_{sam}/E_{ref} \qquad (1)$$

where $E_{sam}$, $E_{ref}$, and t represent the sample, reference, and normalized far-field transmitted electric field amplitude, respectively. The electric field enhancement in the insulating gap is defined as[3]:

$$E_{enhancement} = t \,/\, \text{coverage ratio} \qquad (2)$$

The coverage ratio represents the ratio of the area that light can pass through to the area that light cannot pass through per unit cell (see Figure S1b of the Supporting information S1), as follows:

$$\text{coverage ratio} = \frac{2a_y w}{l_x l_y} \qquad (3)$$

The double deep Q-learning Reward can be set to achieve the best field enhancement and have peaks and dips arranged in an appropriate ratio to accommodate optimum nanogap loop arrays that operate at the specific wireless communication frequency of 0.14 THz while maintaining the best electric field enhancement:

$$E_{enhancement}^{max} = t_{max}/ \text{coverage ratio} \quad (4)$$

$$R_{enhancement} = \log(E_{enhancement}^{max}) \quad (5)$$

$$R_{dist} = (2.5 - |x_{ptarget} - x_{peak}|) \quad (6)$$

$$R_{dip} = (2.5 - (|x_{dtarget} - x_{dip}|) \quad (7)$$

$$\text{Total Reward} = 10 \times R_{enhancement} + R_{dist}^5 + R_{dip}^5 \quad (8)$$

where $R_{enhancement}$ is the Reward for the electric field enhancement, $t_{max}$ is the peak value of the transmitted amplitude spectrum, $R_{dist}$ is the Reward for the peak position in the frequency spectrum, $x_{ptarget}$ is the target peak position, $x_{peak}$ is the first peak position from the frequency spectrum, $R_{dip}$ is the reward for the dip position in the frequency spectrum, $x_{dtarget}$ is the target dip position, $x_{dip}$ is first dip position from the frequency spectrum. A Reward ratio of similar values for field enhancement, peak position, and dip position would not allow convergence to occur. Therefore, we deliberately increased the Reward ratios for peak and dip positions.

The parameter ranges in Figure 2a are $a_x$: 10 ~ 500 μm with 5 μm interval, $a_y$: 10 ~ 500 μm with 5 μm interval, $l_x$: 20 ~ 1000 μm with 5 μm interval, $l_y$: 20 ~ 1000 μm with 5 μm interval, h: 30 ~ 200 nm with 10 nm interval, and w: 2 ~ 20 nm with 2 nm interval. So, the total number of cases is 68,466,061,620. Further, it should be noted that the ranges of the thickness of the thin film as well as the width of the gap can be restricted depending on the fabrication methods used. The target peak position is set to 0.14 THz, one candidate of the 6G communication frequencies [46].

To avoid an overlap with the first Rayleigh minimum, the target dip position is set to 0.4 THz to maintain a sufficient distance from the peak on the spectrum. With our inverse design method, the optimal structure was calculated in 39 hours with the following parameters: $a_x = 10$ μm, $a_y = 200$ μm, $l_x = 215$ μm, $l_y = 625$ μm, $h = 30$ nm, and $w = 2$ nm. The Figure 2b illustrates how the spectrum and Rewards change when the parameters are changed from the optimal structure. For instance, the Reward will be reduced substantially if the dip position changes significantly by $l_x$ while the other parameters remain the same.

In the nanogap loop arrays, we have demonstrated that the strongest field enhancement at 0.14 THz can be realized through the THz nanogap loop with the largest aspect ratio of the rectangular loop, $a_y/a_x=20$, while the perimeter of the loop determines the resonance peak frequency[12, 43], as shown in Figure 3a,b. Interestingly, as we modify the horizontal and vertical periodicities within the given ranges, the field enhancement changes significantly while the peak position remains relatively unchanged, in Figure 3c,d. Lastly, as the nanogap width and the metal thickness simultaneously decrease as shown in Figure 3e,f, the field enhancement can be strongly enhanced in the same manner as in the earlier papers[16, 47]. Consequently, all parameters affect the amplitude and position of the resonant peak significantly. The strong coupling between parameters such as shape, period, thickness, and gap width for THz nanogap loop arrays is illustrated in Figure 3a-f, which makes it difficult to determine the desired structure. Therefore, since all six parameters must be considered to obtain the best THz nanoresonator, it is very suitable to use the inverse design method instead of what is intuitively determined. Despite the fact that rapid inverse design allows for the design of THz nanogap loop arrays by combining modal expansion with double deep Q-learning, For the implementation of next generation communication applications, limitations of the obtained design parameters should be considered in terms of fabrication methods and measurements.

**Realization of inverse designed terahertz nanogap loop arrays.** We employed atomic layer lithography[16] (For the details of the sample fabrication, see the Methods and the Supporting information S2) to fabricate THz nanogap loop structures with a gap width of 2 nm over a large area, in order to achieve the optimum THz nanogap loop arrays obtained by the rapid inverse design method. Figure 4a-c demonstrate that the atomic layer lithography, which combines conventional photolithography and atomic layer deposition, allows us to obtain the THz nanogap loop arrays with exact periodicities, aspect ratios, and the gap width of 2 nm without significant deviations from the optimal design. In Figure 4d, we carried out THz-TDS to obtain THz transmitted electric field spectra through our THz nanogap loop arrays (For the details of the THz-TDS, see the Methods and the Supporting information S3) and estimated the corresponding field enhancement at the nanogap from the far-field measurements using the vector diffraction theory[3]. Interestingly, with the optimal design parameters, we achieved a best field enhancement of about 32,000 at the resonance of 0.2 THz, which is 300 % stronger than those previously achieved at that frequency[16]. It should be noted that the best field enhancement at the 6G communication frequency was quantitatively confirmed by both experimental and numerical methods (For the details of the numerical simulation, see the Methods).

For a better understanding of the stronger field enhancement compared to earlier works, Figure 4e,f show how the electric field enhancement is influenced by the y-period based on the optimized structure, in contrast to many studies using x-polarized incident light which did not consider the y-period effect[8-12, 44, 47]. Indeed, the shortest y-period was preferred to achieve a higher THz transmittance through the nanogap loop array. Interestingly, as the y-period increases from 210 to 625 μm, the coverage ratio decreases even more than the THz transmittance, leading to a higher field enhancement while maintaining the peak frequency relatively unchanged, as shown in Figure 4e,f. Furthermore, we note that the resonance peak position undergoes a blue-shift compared to the desired resonance frequency of 0.14 THz, and there is a notable increase of about 11 % in field enhancement. Figure 5a,b show the transmission electron microscopy (TEM) and the energy-dispersive X-ray spectroscopy (EDS) images of the cross-sectional view of the entire nanogap, showing the slanted gap at the top and widened gap at the bottom of the 2 nm gap. To consider such a structure, we assumed that the widened region is filled with air and carried out a numerical simulation. To avoid excessive computational resources, the full numerical simulations were performed using a three-dimensional finite element method with a reduced y-period of 210 μm. In Figure 5c, when compared to the ideal gap (case I), the slant gap (case II) demonstrated that the resonance peak is blue-shifted in a similar manner to Figure 4d due to the widened region filled with air, and the maximum field enhancement is increased due to the thinner thickness of metal surrounding the 2 nm gap. Figure 5d,e represent the electric field distributions in the ideal and widened gaps, and most of the light is concentrated in the narrowest region of the nanogap, indicating that the majority of the electric field enhancement occurs in the narrowest region of the nanogap.

Conclusion

Previous limitations in the inverse design of terahertz nanophotonic devices, such as the extensive computational time of numerical simulations and the discrepancy between wavelength and structure size, have been overcome. Our proposed rapid inverse design method, incorporating an analytical solution based on the modal expansion method, allows for the efficient attainment of maximum field enhancements in THz nanogap loop arrays within a few days. Experimental results demonstrated an impressive electric field enhancement of approximately 32,000 times at the 2 nm gaps over a large area. While the analytical solution provides a single-mode approximation, slight variations in spectral shape and amplitude may occur in reality. Despite its restriction to specific structures, the analytical solution-based inverse design method significantly outperforms simulation-based approaches in terms of computational speed. Thus, our inverse design method holds immense promise for designing terahertz nano-photonic devices for next-generation communication technologies and molecular sensing applications, serving as a viable alternative to simulation-based inverse design methods. Furthermore, this methodology opens up new possibilities for exploring terahertz nonlinear and quantum phenomena[48, 49].

Methods

**Inverse design method.** Our inverse design method is constructed using the "reinforcement learning designer" toolbox in MATLAB. The Environment consists of the analytical solution from the Supporting information S1 and the Reward obtained from the electric field transmission spectrum. The Agent based on the double deep Q-learning algorithm consists of two layers with 256 nodes per layer. Leaky ReLU and Adam were also employed as an activation function and an optimizer, respectively, and L2 regularization was applied. The epsilon-greedy policy is employed to introduce a stochastic element when the Agent selects an action. In order to design the nanogap loop arrays, it takes 39 hours, and approximately 200,000 calculations are required on a typical middle-level computer (CPU: 3.40 GHz, 6 cores, 12 threads, RAM: 16 GB, GPU: NVIDIA GeForce GTX 1050).

**Finite element method (FEM) simulation.** For the FEM simulation of nanogap loop array, the wave optics module of COMSOL MULTIPHYSICS software was employed. In the geometry, a rectangular hole array structure was created, and periodic boundary conditions were set to the xz- and yz-boundaries. For reducing the computational resource, quarter geometry was employed. The perfect electric conductor and the perfect magnetic conductor conditions were set to xz- and yz-boundaries, respectively. In the simulation of the optimal structure, the structure was composed of a gold film with a nanogap made of $Al_2O_3$. The bottom of the pattern was set to a Si substrate and the top to air. For the slanted nanogap, the slanted region on top and the widened region at the bottom were set to air, and the nanogap region to $Al_2O_3$, as shown in Figure 5e. To prevent unwanted reflections, perfectly matched layers were applied to the top and bottom of all simulations. The refractive indices of air, $Al_2O_3$ and silicon substrate were set to 1.0, 1.73 and 3.4, respectively, and the Drude model was applied to extract the complex refractive indices of gold. For the simulations, we used the workstation with AMD Ryzen Threadripper 3990X (64 cores, 128 threads) and 256 GB RAM. In order to calculate a single state of the nanogap loop array, it took about 20 hours.

**Sample fabrication.** A photoresist pattern was first fabricated on a silicon substrate. The first metal was then deposited on the sample. Following the deposition, the lift-off process was carried out with n-methyl-pyrrolidone solution. An atomic layer deposition process was used to deposit $Al_2O_3$ as an insulating gap material for the nanogap. Using an adhesive tape, the excess metal over the first metal pattern can be peeled off. The schematic diagrams of the entire fabrication process are shown in the Figure S2 of the Supporting information S2.

**Terahertz time-domain spectroscopy (THz-TDS).** An ultrafast optical parametric chirped-pulse amplification (OPCPA) laser[50] was used as a seed laser, which has a center-wavelength of 800 nm, a power of 4 W, a pulse width of 9 fs, and a repetition rate of 1 MHz. The GaAs photoconductive antenna (PCA) was excited by the optical pulse to generate a single-cycle THz pulse, and the THz pulse are guided through the parabolic mirrors to the samples. The transmitted THz pulse was detected by an electro-optic sampling (EO sampling) method using a 1 mm thick ZnTe (110) crystal[16]. The schematic of the THz-TDS and the process for obtaining transmitted amplitude spectra in the frequency domain are shown in Figure S3 of the Supporting information S3.

FIGURES

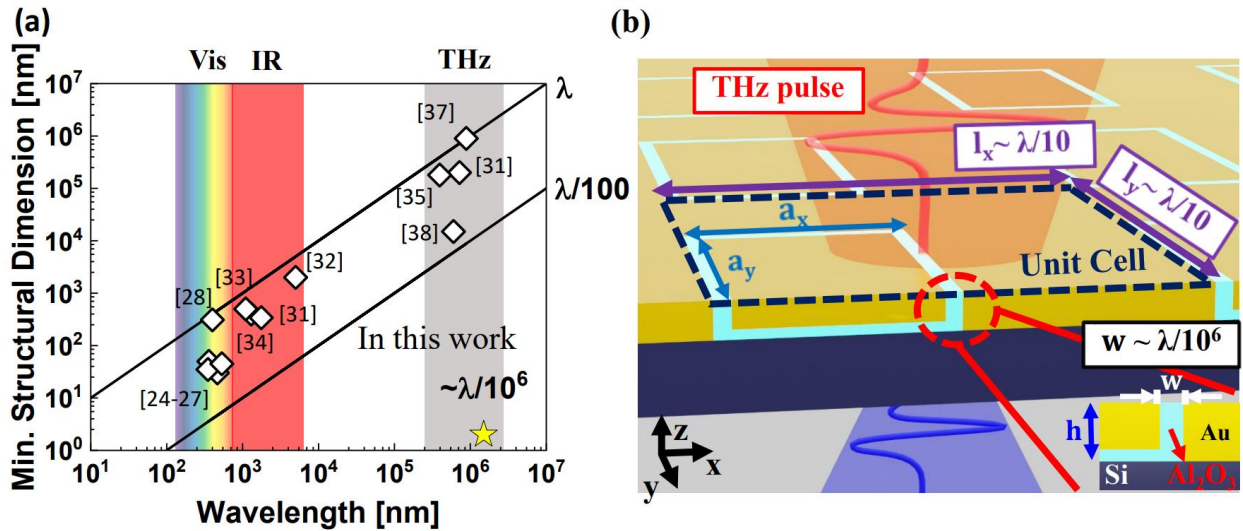

**Figure 1.** (a) According to the reference papers for inverse design methods, a graph is presented for comparing minimum structure dimensions and wavelengths. Our THz nanogap loop arrays operate at millimeter wavelengths while maintaining the minimum gap size of a few nanometers. (b) A schematic of the THz nanogap loop arrays, consisting of $Al_2O_3$ nanogap along the gold rectangular loop on an undoped silicon substrate. Here, $a_x$, $a_y$, $l_x$, $l_y$, h, and w are the width and height of the rectangular loop, the horizontal and vertical period, the thickness of metal, and the width of the nanogap, respectively. For the parameters of $a_x$, $a_y$, $l_x$ and $l_y$, the range is from $\lambda/10$ to $\lambda/100$, while for the parameters of h and w, the range is from $\lambda/1,000,000$ to $\lambda/100,000$.

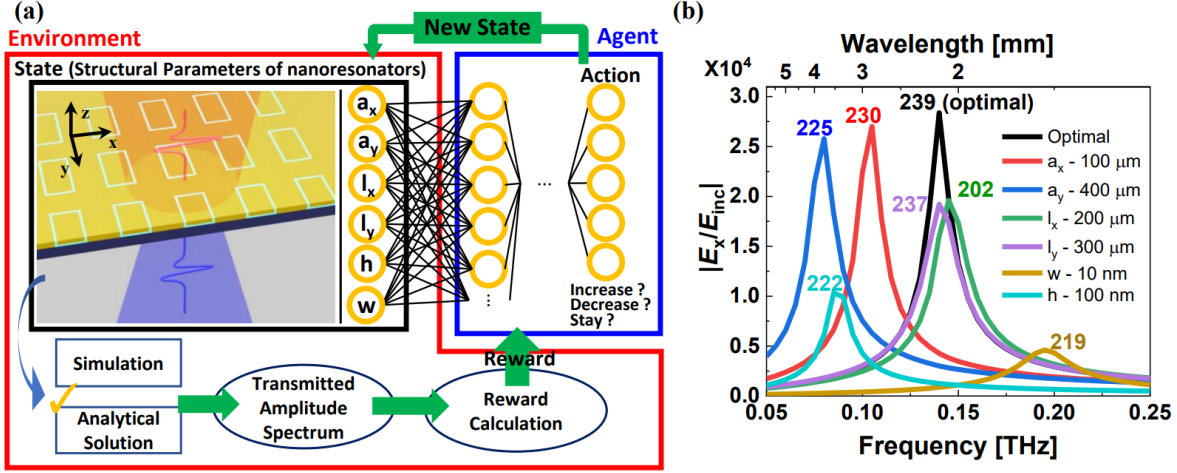

**Figure 2.** (a) Full scheme of an inverse design method using an analytical solution based on the modal expansion. Using the double deep Q-learning method, the Environment consists of a State and a Reward. A State consists of six structural parameters $a_x$, $a_y$, $l_x$, $l_y$, $h$, and $w$ that affect the transmission spectrum. An Action is defined as an increase, decrease, or stay of a parameter. (b) At 0.14 THz, the optimal structure (black line) has the parameters of $a_x$: 10 μm, $a_y$: 200 μm, $l_x$: 215 μm, $l_y$: 625 μm, $h$: 30 nm, and $w$: 2 nm. One specific parameter changes the field enhancement ($|E_x/E_{inc}|$) spectra and Rewards while the other parameters remain constant.

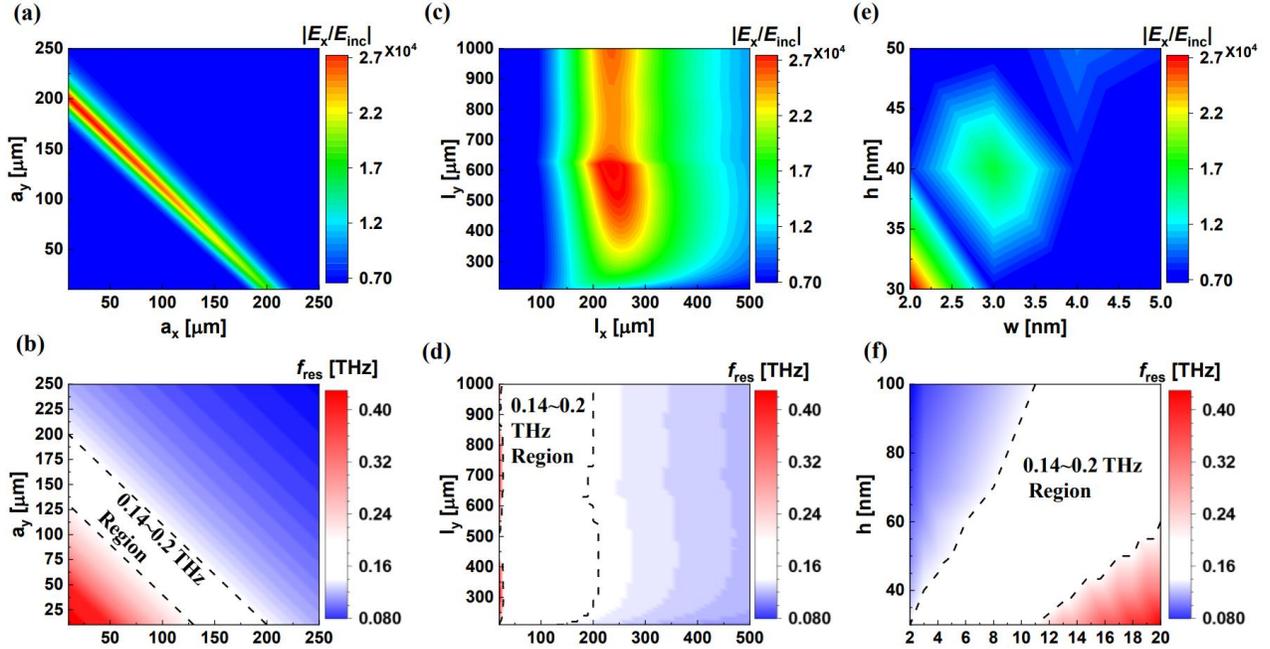

**Figure 3.** The predicted field enhancements ($|E_x/E_{inc}|$ at 0.14 THz) and peak positions ($f_{res}$) by (a, b) the aspect ratio of the rectangular loop with $a_x$ and $a_y$, (c, d) the periodicities with $l_x$ and $l_y$, and (e, f) the metal thicknesses $h$ and nanogap widths $w$.

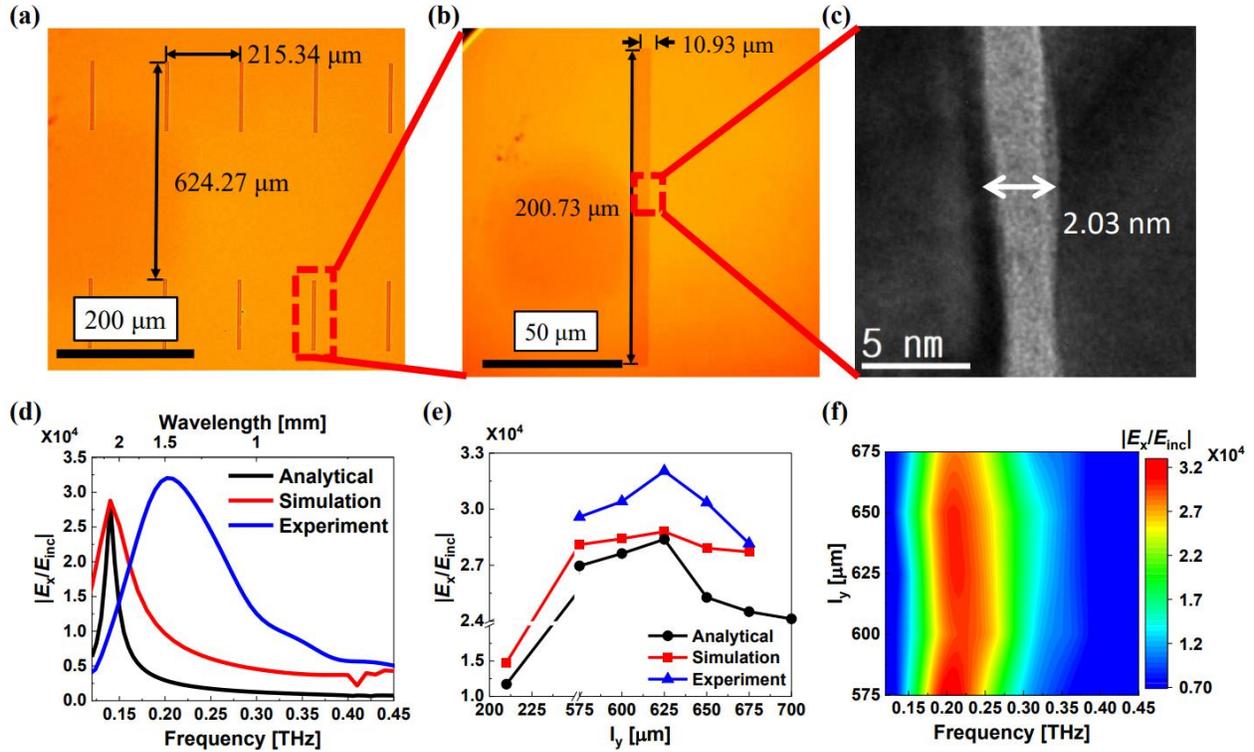

**Figure 4.** (a and b) The microscope images of the optimal THz nanogap loop array. (c) Transmission Electron Microscope (TEM) image for the nanogap area. (d) The field enhancement ($|E_x/E_{inc}|$) spectra of the optimal structure using the analytical, numerical, and experimental methods. (e, f) The peak field enhancements and their spectra, depending on the vertical periodicity of $l_y$.

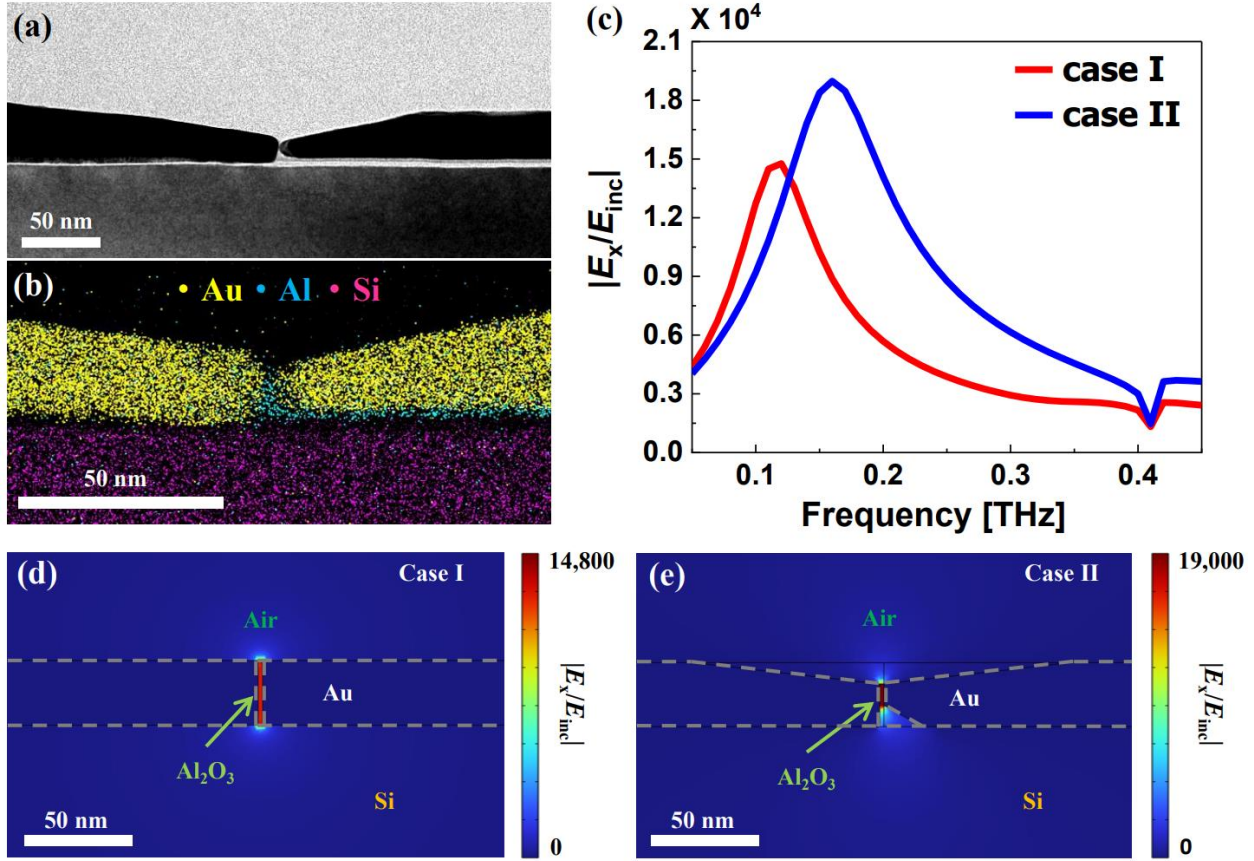

**Figure 5.** (a) Transmission electron microscopy (TEM) (b) Energy-dispersive X-ray spectroscopy (EDS) images of the entire gap region with the narrowest gap width of 2 nm. (c) Comparison of field enhancement ($|E_x/E_{inc}|$) spectra between the ideal (case I, (d)) and slant gap cases (case II, (e)). Electric field distributions for the (case I, (d)) ideal and (case II, (e)) slant gaps are shown.

## ASSOCIATED CONTENT

### Supporting information

Details of the analytical solution for the THz nanogap loop array, fabrication process, terahertz time-domain spectroscopy. (PDF)

## AUTHOR INFORMATION

### Corresponding Author

Hyeong-Ryeol Park - Department of Physics, Ulsan National Institute of Science and

Technology (UNIST), Ulsan 44919, South Korea### Authors

Hyoung-Taek Lee - Department of Physics, Ulsan National Institute of Science and Technology (UNIST), Ulsan 44919, South Korea

Jeonghoon Kim - Department of Physics, Ulsan National Institute of Science and Technology (UNIST), Ulsan 44919, South Korea

Joon Sue Lee - Department of Physics and Astronomy, University of Tennessee, Knoxville, Tennessee 37996, USA

Mina Yoon - Materials Science and Technology Division, Oak Ridge National Laboratory, Oak Ridge, Tennessee 37831, USA

**Author Contributions**

H. -T. L. and H. -R. P. conceived the project. H. -T. L. constructed the inverse design environment and took the simulations. J. K. and H. -T. L. fabricated the samples and conducted the terahertz time-domain spectroscopy experiments. All authors participated the discussions of results and wrote the manuscript together.

**Notes**

The authors declare no competing financial interest.

ACKNOWLEDGMENTS

This work was supported by the National Research Foundation of Korea (NRF) grant funded by the Korean government (NRF-2021R1A2C1008452 and NRF-2022M3H4A1A04096465), the Republic of Korea's MSIT (Ministry of Science and ICT) under the High-Potential Individuals Global Training Program (Task No. 2021-0-01580) and the ITRC (Information Technology Research Center) support program (IITP-2023-RS-2023-00259676) supervised by the IITP (Institute of Information and Communications Technology Planning & Evaluation), and 2023 Research Fund (1.230022.01) of Ulsan National Institute of Science and Technology (UNIST), and also partially was supported by the US Department of Energy, Office of Science, Office of Basic Energy Sciences, Materials Sciences and Engineering Division and by the U.S. DOE, Office of Science, National Quantum Information Science Research Centers, Quantum Science Center (M. Y.).

# Over 30,000-fold field enhancement of terahertz nanoresonators enabled by rapid inverse design

**Hyoung-Taek Lee[1], Jeonghoon Kim[1], Joon Sue Lee[2], Mina Yoon[3], and Hyeong-Ryeol Park[1,*]**

[*]Correspondence: nano@unist.ac.kr
[1]Department of Physics, Ulsan National Institute of Science and Technology (UNIST), Ulsan 44919, South Korea
[2]Department of Physics and Astronomy, University of Tennessee, Knoxville, Tennessee 37996, USA
[3]Materials Science and Technology Division, Oak Ridge National Laboratory, Oak Ridge, Tennessee 37831, USA

# S1. Analytical solution for the THz nanogap loop array

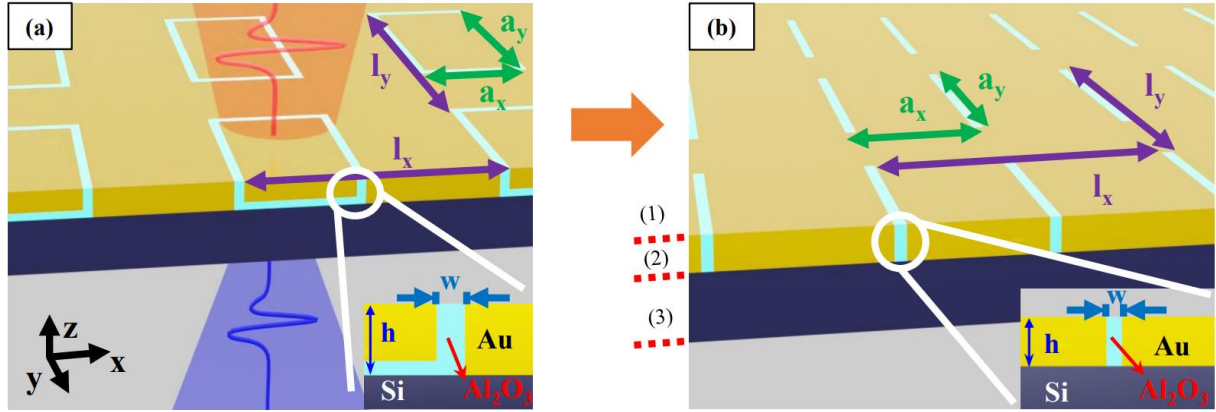

**Figure S1.** Schematic diagrams of (a) the THz nanogap loop and (b) simplified two nano-slots for the analytical calculation based on modal expansion method.

In order to get the analytical solution for the THz nanogap loop array, it is necessary to simplify the nanogap loop to simplified two nano-slots, as shown in Fig. S1. When the light with an x-polarized electric field is incident on the nanogap loop array, the electric field in the gap is [1]:

$$E_{gap} = \frac{G_V I_0}{(G_1 - G_V \cos \beta h)(G_3 - G_V \cos \beta h) - G_V^2}$$

From the boundary conditions, each symbol represents the following:

$$G_{1,3} = \frac{i(a_x + a_y)w}{2(l_x + l_y)} \left( \frac{\varepsilon_{1,3} k_0 (k_0 + Z_s k_{1z,3z}) - k_y^2}{(k_0 + Z_s k_{1z,3z})(k_{1z,3z} + Z_s k_0)} \right) J^2$$

where $a_x$ and $a_y$ are the width and height of the rectangular loop, $l_x$ and $l_y$ are the horizontal and vertical periods, $w$ represents the width of the nanogap, $\varepsilon_{1,3}$ are the dielectric function of area 1 (air) and area 3 (substrate), $k_0$ and $Z_s$ are the wavevector and the impedance in the vacuum, respectively, as shown in Fig. S1b.

$$J = \mathrm{sinc}(k_x w/2) \left\{ \mathrm{sinc}\left(\frac{\pi + k_y(a_x + a_y)}{2}\right) + \mathrm{sinc}\left(\frac{\pi - k_y(a_x + a_y)}{2}\right) \right\}$$

$$k_{1z,3z} = \sqrt{\varepsilon_{1,3} k_0^2 - (k_x^2 + k_y^2)}$$

where $k_x$ and $k_y$ are summation of the integer multiples of $2\pi/a_x$ and $2\pi/a_y$, respectively.

when the gap size is reduced to the nanometer scale of below 10 nm, the gap plasmon effect should be considered [2]. Due to the gap plasmon effect, the wavevector leaks into the metal. This is considered as follows for the dielectric function of the metal:

$$G_V = \frac{1}{k_0 h \left(1 + \frac{k_m^2}{\epsilon_d k_0^2}\right) \times \mathrm{sinc}(\beta h)}, \quad \beta = \sqrt{\epsilon_d k_0^2 + k_m^2 - \left(\frac{\pi}{a_x + a_y}\right)^2}$$

where h is the thickness of the metal film, $\epsilon_d$ is dielectric function of insulator (gap material), and the leaking wavevector $k_m$ is defined as:

$$k_m^2 = -\frac{2\left(1+\sqrt{1-k_0^2\left(\frac{w\epsilon_m}{\epsilon_d}\right)^2\times(\epsilon_m-\epsilon_d)}\right)}{\left(\frac{w\epsilon_m}{\epsilon_d}\right)^2}$$

where $\epsilon_m$ is dielectric function of metal. Outside the gap, the electric field is expressed as:

$$E_{out} = \frac{E_{gap}(n_1+n_3)(a_x+a_y)}{2a_x\pi}\int_{\frac{\pi}{2}-\frac{a_x}{a_x+a_y}}^{\frac{\pi}{2}+\frac{a_x}{a_x+a_y}} \cos x \, dx$$

where $n_{1,3}$ are refractive index of area 1 (air) and area 3 (substrate), respectively, as shown in Fig. S1b. Accordingly, the transmitted electric field amplitude can be obtained after normalizing by the ratio between the areas of the dielectric and metal per unit cell.

$$t = E_{out}\frac{w(a_x+a_y)}{l_x l_y}$$

## S2. Fabrication process of the THz nanogap loop arrays

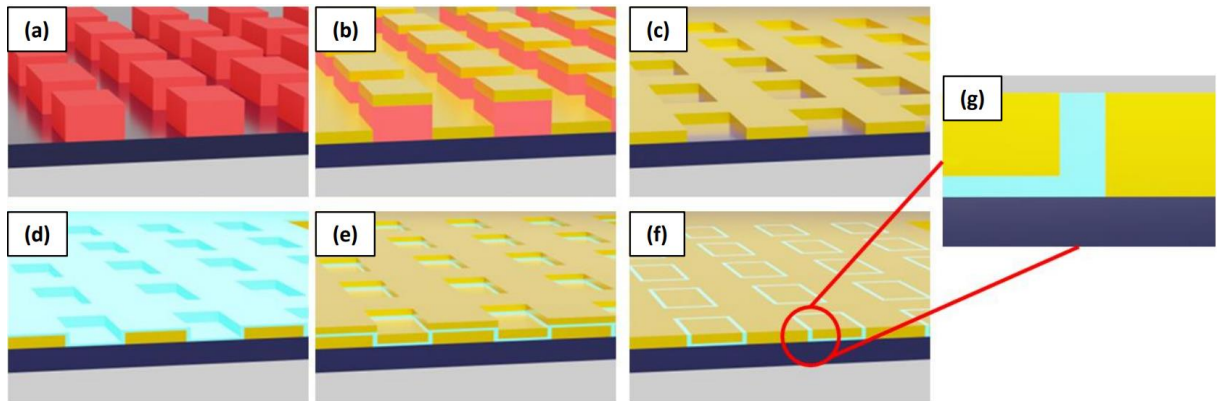

**Figure S2.** The fabrication process of the nanogap loop array structure by atomic layer lithography [2]. (a) The photoresist patterns are fabricated on the undoped silicon substrate. (b) The first metal is deposited, and (c) the photoresist is removed by n-methyl- pyrrolidone (NMP) solution. (d) The gap material ($Al_2O_3$) is deposited by the atomic layer deposition (ALD) with sub-nanometer precision. (e) The second metal is deposited, and (f) the excess metal is peeled-off mechanically by an adhesive tape. (g) The cross-section view of the nano-gap structure.

## S3. Terahertz time-domain spectroscopy (THz-TDS)

(a)

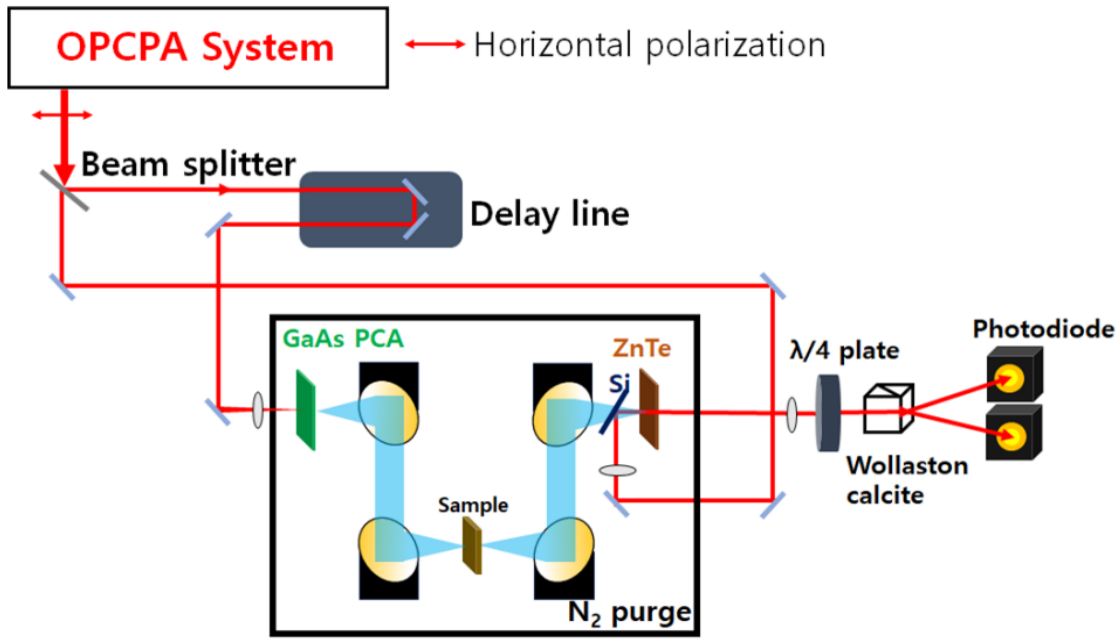

(b)

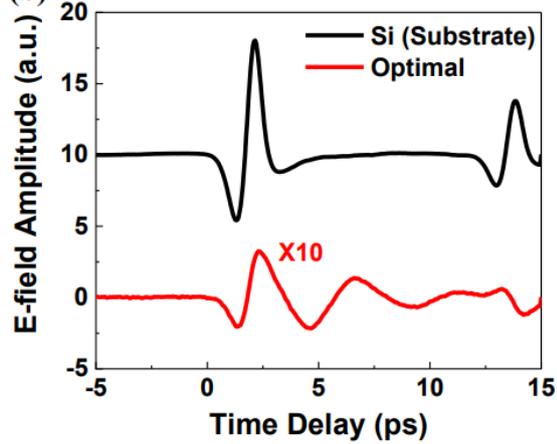

(c)

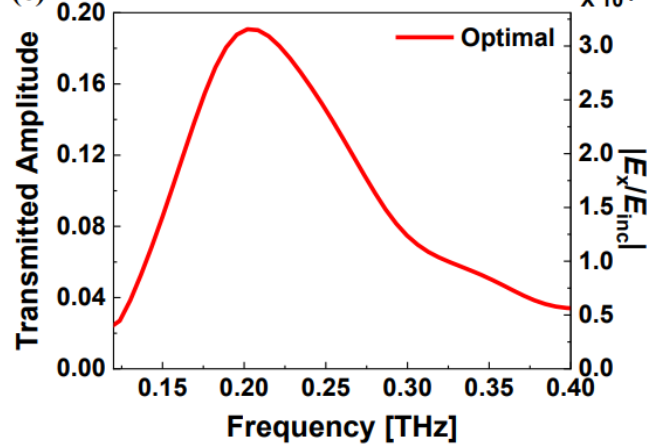

**Figure S3.** (a) Schematic of terahertz time-domain spectroscopy (THz-TDS). (b) Measured THz time traces for the silicon substrate as a reference (black line) and the optimal nanogap loop array (red line) using THz-TDS. (c) After Fourier-transforming the time traces in (b), the transmitted amplitude of the optimal nanogap loop array can be converted to the field enhancement ($|E_x/E_{inc}|$) by the vector diffraction theory [3]. This transmitted amplitude is normalized by the reference spectrum with the silicon substrate.